\documentclass[11pt]{article}

\newtheorem{lemma}{Lemma}
\newtheorem{proposition}{Proposition}
\newtheorem{theorem}{Theorem}
\newtheorem{corollary}{Corollary}

\def\tr{\mathop{\rm tr}\nolimits}

\def\dif{\mathop{\rm d}\nolimits}
\def\ci{\mathop{\textrm{i}}\nolimits}

\begin{document}

\title{On the space-times admitting two shear-free geodesic null congruences}
\author{Joan Josep Ferrando$^1$\ and Juan Antonio S\'aez$^2$}

\maketitle \vspace{1cm}

\begin{abstract}
We analyze the space-times admitting two shear-free geodesic null
congruences. The integrability conditions are presented in a plain
tensorial way as equations on the volume element $U$ of the
time-like 2--plane that these directions define. From these we
easily deduce significant consequences. We obtain explicit
expressions for the Ricci and Weyl tensors in terms of $U$ and its
first and second order covariant derivatives. We study the different
compatible Petrov-Bel types and give the necessary and sufficient
conditions that characterize every type in terms of $U$. The type D
case is analyzed in detail and we show that every type D space-time
admitting a 2+2 conformal Killing tensor also admits a conformal
Killing-Yano tensor.
\end{abstract}
\section{Introduction}
\label{intro}

The family of space-times admitting two shear-free geodesic null
congruences contains significant and well-known solutions of the
Einstein equations. Hence, in the Schwarzschild and Kerr black holes
these congruences define the outgoing and ingoing propagation of
light in the radial direction. We find a similar situation in their
charged counterparts, the Reissner-Nordstr\"om and Kerr-Newman
solutions. Moreover, all the other vacuum Petrov-Bel type D
solutions and their charged counterparts also belong to this family.

But this family contains other interesting space-times. Indeed, we
have shown elsewhere \cite{fsD} that in every type D metric with a
vanishing Cotton tensor the two principal planes define an umbilical
2+2 structure, and that this geometric property states,
equivalently, that the two principal null directions determine
shear-free geodesic congruences. It is worth remarking that the
umbilical condition can be stated in terms of the {\em canonical
2--form} $U$, volume element of the time-like plane defined by the
two congruences \cite{fsD}.

On the other hand, some first integrals of the geodesic equation are
closely linked with this property. Thus, conformal and Killing
tensors of type 2+2 and non-null Killing-Yano tensors can only exist
in space-times admitting two shear-free geodesic null congruences
\cite{hm-1,diru-1,fsY,cfsKT}.

A shear-free geodesic null congruence determines a Debever principal
null direction of the Weyl tensor, but is not, necessarily, a
multiple Debever direction. Consequently, all the Petrov-Bel types,
excepting type N, are compatible with an umbilical 2+2 structure.
What restrictions does this condition impose on the Weyl and Ricci
tensors? In order to give an answer to this question one needs to
study the integrability conditions of the umbilical equation.

These integrability conditions were presented in spinorial formalism
by Dietz and R\"{u}diger \cite{diru-1} and they obtained some
complementary restrictions on the type D space-times. Here, we
develop this question in a plain tensorial way that allows us to
analyze, not only all the compatible Petrov-Bel types, but also the
restrictions on the Ricci tensor.

Our results can easily be summarized if one takes into account that
when $U$ defines an umbilical structure, its first derivatives can
be collected in a complex vector $\chi$ \cite{fsD,fsY}. Then the
covariant derivative $\nabla \chi$ contains all the information on
the second derivatives of $U$. We show that the Ricci tensor is
determined, up to two functions, by $U$, $\chi$ and the symmetric
part of $\nabla \chi$, and we give its explicit expression. On the
other hand we give a general expression for the Weyl tensor in terms
of the $U$, $\chi$, $\dif \chi$ and the scalar curvature. Moreover
we study the characterization of the different Petrov-Bel types and
we show that the 2--forms $U$ and $\dif \chi$ determine the
canonical 2--forms associated with the Weyl tensor geometry.

From our study a result easily follows: if a non conformally flat
space-time admits a non-null Maxwell field whose principal
directions define shear-free geodesic null congruences, then the
metric is type D and these directions are the double Deveber
directions of the Weyl tensor. This property has allowed elsewhere
\cite{fsY} to obtain the intrinsic characterization of the
space-times admitting Killing-Yano or conformal Killing-Yano
tensors, and to give an algorithm that determines these first
integrals of the geodesic equation.

It is known that a type D vacuum solution admitting a Killing tensor
also admits a Killing-Yano tensor \cite{collinson,ste}. On the other
hand, all the type D vacuum solutions admit a 2+2 conformal Killing
tensor \cite{walker-p} and, from our results here and \cite{fsY},
they also admit a conformal Killing-Yano tensor. Can these results
be generalized to the non vacuum case? Here we show that {\em every
type D solution with a 2+2 conformal Killing tensor also admits a
conformal Killing-Yano tensor}.

This paper is organized as follows. In section 2 we introduce the
notation and summarize some previous results needed in this work. In
section 3 we present the integrability conditions  for the umbilical
equation and we obtain the restrictions that these conditions impose
on the Ricci and Weyl tensors. The compatible Petrov-Bel types are
considered in section 4 and we show that the Weyl canonical
bivectors are determined by $U$ and its differential concomitants.
Finally, section 5 is devoted to analyzing type D space-times in
detail and we recover some known result and provide new ones.

\section{Notation and previous results}
\label{sec-notation}

Let $(V_4,g)$ be an oriented space-time of signature $\{ -, +,+,+
\}$. The metric $G$ on the space of 2--forms is $G=\frac{1}{2} g
\wedge g$, $\wedge$ denoting the double-forms exterior product, $(A
\wedge B)_{\alpha \beta \mu \nu} = A_{\alpha \mu} B_{\beta \nu} +
A_{\beta \nu} B_{\alpha \mu} - A_{\alpha \nu} B_{\beta \mu} -
A_{\beta \mu} B_{\alpha \nu}$. If $F$ and $H$ are 2--forms, $(F,H)$
denotes the product with $G$:
$$
(F,H) \equiv G(F,H) = \frac14 G_{\alpha \beta \lambda \mu} F^{\alpha
\beta} H^{\lambda \mu} = \frac12 F_{\alpha \beta} H^{\alpha \beta}\,
,
$$

A self--dual 2--form is a complex 2--form ${\cal F}$ such that
$*{\cal F}= \textrm{i}{\cal F}$, where $*$ is the Hodge dual
operator. We can associate biunivocally with every real 2--form $F$
the self-dual 2--form ${\cal F}=\frac{1}{\sqrt{2}}(F-\textrm{i}*F)$.
For short, we here refer to a self--dual 2--form as a {\it SD
bivector}. The endowed metric on the 3-dimensional complex space of
the SD bivectors is ${\cal G}=\frac{1}{2}(G-\textrm{i} \; \eta)$,
$\eta$ being the metric volume element of the space-time. The
orthogonal complement of the SD bivectors space is the space of the
anti-self-dual 2--forms, which are those satisfying $*F = -
\textrm{i}F$.

If $F$ is a 2--form and $P$ and $Q$ are double-2--forms, $P(F)$ and
$P \circ Q$ denote, respectively, the 2--form and the double-2--form
given by:
$$
P(F)_{\alpha \beta} \equiv \frac12 P_{\alpha \beta}^{\ \ \mu \nu}
F_{\mu \nu}, \qquad (P \circ Q)_{\alpha \beta \rho \sigma} \equiv
\frac12 P_{\alpha \beta}^{\ \ \mu \nu} Q_{\mu \nu \rho \sigma}
$$

Every double-2--form, and in particular the Weyl tensor $W$, can be
considered as an endomorphism on the space of the 2--forms. The
restriction of the Weyl tensor on the SD bivectors space is the {\em
self-dual Weyl tensor} and it is given by:
$$
{\cal W} \equiv {\cal G} \circ W \circ {\cal G} = \frac12(W - \ci
*W)
$$

For short we here refer to a p-dimensional distribution (set of
vector fields generated by p independent vector fields) as a
p--plane. The generalized second fundamental form $Q_v$ of a
non-null p--plane $V$ is the (2,1)-tensor:
\begin{equation}
Q_v(x,y) = h(\nabla_{v(x)}v(y)) \, , \qquad \forall \; \; x,y
\end{equation}
where $v$ is the projector on $V$. Let us consider the invariant
decomposition of $Q_v$ into its antisymmetric part $A_v$ and its
symmetric part $S_v \equiv S_v^T + {1 \over p}v \otimes \tr S_v$,
where $S_v^T$ is a traceless tensor:
\begin{equation}  \label{Q2}
Q_v = A_v + \frac{1}{p} v \otimes \tr S_v + S_v^T
\end{equation}
The p-plane $V$ is a {\em foliation} if, and only if, $A_v =0$, and,
similarly, $V$ is said to be {\em minimal, umbilical or geodesic} if
$\tr S_v=0$, $S_v^T =0$ or $S_v  =0$, respectively.

A p+q almost-product structure is defined by a p-plane $V$ and its
orthogonal complement q-plane $H$. The almost-product structures can
be classified by taking into account the invariant decomposition of
the covariant derivative of the so called structure tensor $\Pi$
\cite{nav} or, equivalently, according to the foliation, minimal or
umbilical character of each plane \cite{fsD,olga}. We will say that
a structure is integrable when both, $V$ and $H$ are a foliation. We
will say that the structure is minimal (umbilical) if both, $V$ and
$H$ are minimal (umbilical).

On the space-time, a 2+2 almost-product structure is defined by a
time-like plane $V$ and its space-like orthogonal complement $H$.
Let $v$ and $h= g-v$ be the respective projectors and let $\Pi =
v-h$ be the {\it structure tensor}. A 2+2 space-time structure is
also determined by the {\it canonical} unitary 2-form $U$, volume
element of the time-like plane $V$. Then, the respective projectors
are $v=U^2$ and $h = -(*U)^2$, where $U^2 = U \cdot U$. Here, if $A$
and $B$ are two 2--tensors, we denote $A \cdot B$ the tensor with
components $(A \cdot B)_{\alpha \beta} = A_{\alpha}^{\ \mu} B_{\mu
\beta}$.

In working with 2+2 structures it is useful to introduce the
canonical SD bivector ${\cal U} \equiv \frac{1}{\sqrt{2}} (U - {\rm
{i}} *U )$ associated with $U$, that satisfies $2{\cal U}^2 = g$
and, consequently, it is unitary, $({\cal U},{\cal U}) = -1$. If
$\perp$ denotes the projector on the space of the SD bivectors which
are orthogonal to ${\cal U}$, and ${\cal G}_{\perp}$ is the
restriction of the metric on this space, for a 2--form $F$ and a
double-2--form $P$, we have:
\begin{equation}
{\cal G}_{\perp} = {\cal U} \otimes {\cal U} + {\cal G} \, , \qquad
F_{\perp} \equiv {\cal G}_{\perp}(F) \, , \qquad  P_{\perp} \equiv
{\cal G}_{\perp} \circ P \circ {\cal G}_{\perp}
\end{equation}

If $A$ and $B$ are two 2--tensors, $[A,B]$ and $\{A,B\}$ denote the
commutator and anti-commutator, respectively:
$$
[A,B] = A \cdot B - B \cdot A  \, ,\qquad \{A,B\} = A \cdot B + B
\cdot A
$$
With this notation a straightforward calculation leads to:
\begin{lemma} \label{lemma-real-U}
A real symmetric tensor $E$ is determined, up to two scalars $e_1$
and $e_2$, by the commutator $[{\cal U}, E]$, where ${\cal U}$ is a
unitary SD bivector. More precisely:
\begin{equation} \label{real-U}
E = \frac{1}{4}(e_1 \, g + e_2 \, \Pi) + {\cal U} \cdot \Big( [{\cal
U} , E] +  \{ {\cal U} , \bar{\cal U} \cdot [\overline{{\cal U} , E}
] \} \Big)
\end{equation}
where $\Pi$ is the structure tensor associated with ${\cal U}$,
$\Pi= 2{\cal U} \cdot \bar{{\cal U}}$, and $\bar{\ }$ stands for
complex conjugate.
\end{lemma}

There are some first order differential concomitants of $U$ that
determine the geometric properties of the structure \cite{fsD}.
Moreover, the first order differential properties of a 2+2 structure
admit a kinematical interpretation \cite{cf}. Now we summarize some
of these results needed in the following sections. If $i_{\xi}$
denotes the interior product with a vector field $\xi$, and $\delta$
the exterior codifferential, $\delta = *d*$, we have the following
lemma \cite{fsD}:

\begin{lemma}  \label{lem-umb}
Let us consider the 2+2 structure defined by the canonical
2--form $U$. The three following conditions are equivalent: \\
(i) The structure is umbilical\\
(ii) The canonical SD bivector ${\cal U}=\frac{1}{\sqrt{2}} (U -
{\rm i} *U)$ satisfies:
\begin{equation}
\Sigma[U] \equiv \nabla {\cal U} - i_{\xi} {\cal U} \otimes {\cal U}
- i_{\xi}{\cal G}=0 \, , \qquad \xi \equiv \delta {\cal U}
\label{umb}
\end{equation}
(iii) The principal directions of $U$ determine shear-free geodesic
null congruences.
\end{lemma}
In the following we refer to (\ref{umb}) as the {\em umbilical
equation}.

On the other hand, let us define the 1--forms:
\begin{equation} \label{Phi-Psi}
\begin{array}{l}
\Phi  \equiv \Phi[U] \equiv *U(\delta*U) - U(\delta U)  \\[1mm]
\Psi \equiv \Psi[U] \equiv *U(\delta U) + U(\delta *U)
\end{array}
\end{equation}
where, for a 2-tensor $A$ and a vector $x$, $A(x)_{\mu} =  A_{\mu
\nu} x^{\nu}$. Then, we have the following result \cite{fsD}:
\begin{lemma} \label{lem-m-i}
Let $U$ be the canonical 2--form of a 2+2 almost product
structure. Then it holds:\\
(i) The structure is minimal if, and only if, $U$ satisfies
$\, \Phi[U]=0$.\\
(ii) The structure is integrable if, and only if, $U$ satisfies $\,
\Psi[U] =0$.
\end{lemma}

These characterizations of a minimal and an integrable structure,
and the kinematic interpretation of these geometric properties given
in \cite{cf} lead to call the $\Phi$ and $\Psi$ given in
(\ref{Phi-Psi}) the {\em expansion vector} and the {\em rotation
vector} of the structure, respectively.

Every non-null 2-form $F$ can be written in the form $F =
e^{\phi}[\cos \psi U + \sin \psi *U]$, where the {\it geometry} $U$
is a unitary and simple 2--form that determines the 2+2 associated
structure (principal planes), $\phi$ is the {\it energetic index}
and $\psi$ is the {\it Rainich index}. When $F$ is solution of the
source-free Maxwell equations, $\delta F =0$, $\delta *F =0$, one
says that its intrinsic geometry $U$ defines a {\it Maxwellian
structure}. In terms of the intrinsic elements $(U,\phi,\psi)$,
Maxwell equations become \cite{rai,cff}:
\begin{equation} \label{maxwell-rainich}
\dif \phi = \Phi[U] \, , \qquad \qquad   \dif \psi = \Psi[U]
\end{equation}
Then, from (\ref{maxwell-rainich}) the Rainich theorem \cite{rai}
follows:
\begin{lemma}
A unitary 2-form $U$ defines a Maxwellian structure if, and only if,
the expansion and the rotation are closed 1--forms, namely $U$
satisfies:
\begin{equation}\label{max2}
\mbox{\rm d} \Phi[U]  = 0 \, , \qquad  \qquad \mbox{\rm d} \Psi[U] =
0
\end{equation}
\end{lemma}

When the Maxwell-Minkowski energy tensor $T$ associated with a
non-null 2--form is divergence--free, the underlying 2+2 structure
is said to be {\em pre-Maxwellian} \cite{debever}. The conservation
of $T$ is equivalent to the first of the Maxwell-Rainich equations
(\ref{maxwell-rainich}) \cite{fsY}. Consequently: {\em a 2+2
structure is pre-Maxweelian if, and only if, the canonical 2--form
satisfies the first equation in (\ref{max2}).}

We can collect the expansion vector $\Phi$ and the rotation vector
$\Psi$ in a complex vector with a simple expression in terms of the
canonical SD bivector ${\cal U}$. Indeed, if we define $\chi \equiv
\chi[U] = i_{\xi} {\cal U}$, $\xi \equiv \delta{\cal U}$, we have:
\begin{equation} \label{chi}
\chi = \frac12 (\Phi[U] + \ci \Psi[U])
\end{equation}
Then, conditions (\ref{max2}) that characterize a Maxwellian
structure can be written as:
\begin{equation} \label{maxwellian-chi}
\dif \chi = 0
\end{equation}
It is worth remarking that in (\ref{maxwellian-chi}) there are just
five independent complex equations (or in (\ref{max2}) ten real
ones). This fact has been pointed out in \cite{cff} and is a
consequence of the integrability condition $\delta \delta {\cal U} =
0$. Indeed this identity states, equivalently, that the complex
2--form $\dif \chi$ is orthogonal to ${\cal U}$:
\begin{equation} \label{max-5}
(\dif \chi, {\cal U}) = 0
\end{equation}
condition that in real formalism becomes:
$$( \dif \Phi, U ) + ( \dif \Psi ,*U) =0 \label{r-1} \, , \qquad
 (\dif \Phi , *U) - (\dif \Psi, U) =0 $$

\section{Integrability conditions for umbilical 2+2-structures}
\label{sec-integrability}

As a consequence of lemma \ref{lem-umb} the Riemann tensor of the
space-times admitting two shear-free geodesic null congruences must
be submitted to the integrability conditions of the umbilical
equation (\ref{umb}). We can obtain these conditions from the Ricci
identities for the SD bivector ${\cal U}$:
\begin{equation} \label{u1}
\nabla_{\alpha}  \nabla_{\beta} \,  {\cal U}_{\mu \nu} -
\nabla_{\beta} \nabla_{\alpha} \,  {\cal U}_{\mu \nu} = {\cal
U}_{\mu}^{\ \lambda} \, R_{\lambda \nu \beta \alpha} - {\cal
U}_{\nu}^{\ \lambda} \, R_{\lambda \mu \beta \alpha}
\end{equation}

The umbilical equation implies that the covariant derivative of the
canonical 2--form ${\cal U}$ is determined by the complex vector
$\xi \equiv \delta {\cal U}$. Indeed, (\ref{umb}) may equivalently
be written as
\begin{equation} \label{u2}
\nabla {\cal U} = i_{\xi} {\cal G}_{\perp} \, , \qquad \xi \equiv
\delta {\cal U} \, .
\end{equation}

If one simplifies (\ref{u1}) by using (\ref{u2}), one obtains:
\begin{proposition} \label{pro-integra-con}
Under the umbilical condition {\rm (\ref{u2})} the Ricci identities
{\rm (\ref{u1})} take the form:
\begin{equation} \label{integra-con-1}
(Riem - K \wedge g) \circ {\cal G}_{\perp} = 0
\end{equation}
\begin{equation} \label{T-xi-chi}
K \equiv T \cdot {\cal U} \, , \qquad    T \equiv \nabla \xi - \chi
\otimes \xi
\end{equation}
where  $\xi = \delta {\cal U}$ and  $\chi = i_{\xi}{\cal U}$.
\end{proposition}

Taking into account  the usual decomposition of the Riemann tensor,
we can write (\ref{integra-con-1}) as equations for the Weyl and
Ricci tensors. Indeed, by considering the self-dual and the
anti-self-dual parts in the first pair of indexes of this equation,
we obtain:
\begin{theorem} \label{theo-integra-con-2}
Let ${\cal W}$, $Ric$ and $R$ be the self-dual Weyl tensor, the
Ricci tensor and the scalar curvature of a space-time admitting an
umbilical structure ${\cal U}$. The integrability conditions for the
umbilical equation {\rm (\ref{umb})} may be written as:
\begin{eqnarray}
{\cal W}_{\perp} &  =  & \Omega {\cal G}_{\perp} \, , \qquad
\Omega \equiv - \left[\frac{R}{12} + (A, {\cal U} )\right]  \label{i-c-1} \\
{\cal W}({\cal U})_{\perp} & = & -A_{\perp}   \label{i-c-2}\\[2mm]
[{\cal U}, Ric] & = & 2 \, {\cal U} \cdot S \cdot {\cal U} - S
\label{i-c-3}
\end{eqnarray}
where $A$ and $S$ are the antisymmetric and symmetric parts of the
tensor $T$ given in {\rm (\ref{T-xi-chi})}:
$$A= \frac{1}{2} (T - T^{t}), \qquad S=\frac{1}{2} ( T + T^{t} )$$
\end{theorem}

The integrability conditions (\ref{i-c-1}) and (\ref{i-c-2}) can be
used to obtain an expression for the Weyl tensor whereas the third
one (\ref{i-c-3}) offers an expression for the Ricci tensor. Below
we give explicit expressions of these two irreducible parts of the
Riemann tensor, and we show that they are determined up to two
scalars by $U$ and its derivatives. This fact is a consequence of
the Codazzi relations, which give all the mixed components of the
curvature tensor as a concomitant of the second fundamental forms of
an arbitrary $p+q$ structure. Thus, only the total projections of
the curvature tensor on the two planes are not determined by the
second fundamental forms. In the case of a 2+2 structure these
projections have one sole component. Thus, if $v(t)$ and $h(t)$
denote the total projection of a tensor $t$ on the planes $V$ and
$H$, respectively, we have:
\begin{equation} \label{curvaturas}
v(Riem) = - X \, U \otimes U , \qquad h (Riem) = Y \,  *U \otimes *U
\end{equation}
Let us note that the two scalars $X$ and $Y$ are the Gauss
curvatures of the respective 2--plane when the structure is a
product one, that is, when the space-time metric breaks down into
two bi-dimensional metrics, $g = v_{ij}(x^k) \dif x^i \dif x^j +
h_{AB}(x^C) \dif x^A \dif x^B$, $i,j,k = 0,1$ and $A,B,C= 2,3$.

Before obtaining the Ricci and Weyl tensors, we study some
integrability restrictions on the structure $U$ that are independent
of the curvature tensor. The third integrability equation
(\ref{i-c-3}) can be written in the form:
\begin{equation} \label{def-P}
[{\cal U}, Ric ] = [{\cal U} , L] \, , \qquad  L \equiv [S , {\cal
U}]
\end{equation}
Thus, the tensor $L$ determines the commutator $[{\cal U}, Ric ]$.
But lemma \ref{lemma-real-U} implies that this commutator fixes the
Ricci tensor up to two scalars. More precisely, considering that
(\ref{i-c-3}) implies $\bar{{\cal U}} \times [\bar{{\cal U}} , Ric ]
= \bar{L}$, and taking $E = Ric$ in expression (\ref{real-U}), we
have:
\begin{equation} \label{ricci-1}
\begin{array}{lll}
Ric & = & \frac{1}{4}( R \ g + r \Pi) + {\cal U} \times [{\cal U} ,
L] + {\cal U} \times \{ {\cal U},
\bar{L} \} = \\[2mm]
& = & \frac{1}{4}( R \ g + r \Pi) + \frac{1}{2} ( L + \bar{L}) +
{\cal U} (L - \bar{L} ) {\cal U}
\end{array}
\end{equation}

The Ricci tensor being real, the imaginary part of the last term in
the above expression must vanish. Then, one obtains:
\begin{equation} \label{integra-con-3-a}
v  (\mbox{\rm{Im}}[ L]) =0 , \qquad h ( \mbox{\rm{Im}} [L]) = 0
\end{equation}

A straightforward calculation allows us to obtain its real and
imaginary parts from the definition (\ref{def-P}) of $L$:
\begin{eqnarray} \label{re-P}
4 \mbox{\rm{Re}}[ L] & = & M - U \cdot M \cdot U + *U \cdot M
\cdot *U - U \cdot N \cdot *U - *U \cdot N \cdot U \\
4 \mbox{\rm{Im}}[ L] & = & N - U \cdot N \cdot U + *U \cdot N \cdot
*U + U \cdot M \cdot *U + *U \cdot M \cdot U  \label{im-P}
\end{eqnarray}
$M$ and $N$ being the symmetric tensors:
\begin{equation} \label{M-N}
M \equiv {\cal L}_{\Phi} g - \Phi \otimes \Phi + \Psi \otimes \Psi
\, , \qquad N \equiv {\cal L}_{\Psi} g - \Phi
\stackrel{\sim}{\otimes} \Psi
\end{equation}
where ${\cal L}_{s}$ denotes the Lie derivative with respect to a
vector field $s$ and, for two arbitrary tensors, $A
\stackrel{\sim}{\otimes} B = A \otimes B + B \otimes A$. We can make
the integrability conditions (\ref{integra-con-3-a}) more explicit.
Indeed, by using (\ref{im-P}) we can easily write these equations
as:
$$
[v(N), U] = 0 \, , \qquad [h[N], *U] = 0
$$
But, taking into account that the sole symmetric tensor in $V$
(resp. $H$) which commutes with $U$ (resp. $*U$) is proportional to
$v$ (resp. $h$), we have:
\begin{proposition} \label{pro-integra-con-3}
The canonical 2--form $U$ of an umbilical structure satisfies the
integrability conditions:
\begin{equation}
v ( \  {\cal L}_{\Psi} g - \Phi \stackrel{\sim}{\otimes} \Psi \ ) =
\lambda \  v \label{r3} \, , \qquad  h ( \  {\cal L}_{\Psi} g - \Phi
\stackrel{\sim}{\otimes} \Psi  \ ) = \mu \  h
\label{integra-con-3-b}
\end{equation}
where $\Phi$ and $\Psi$ are given in {\rm (\ref{Phi-Psi})}, and
$\lambda$ and $\mu$ are two scalars that can be obtained by taking
the trace.
\end{proposition}

\subsection{The Ricci tensor}

Taking its real part, the expression (\ref{ricci-1}) of the Ricci
tensor becomes:
\begin{equation} \label{ricci-2}
Ric = \frac{1}{4}( R \ g + r \Pi) + \mbox{\rm{Re}}[L] -  U \cdot
\mbox{\rm{Im}}[ L ] \cdot *U -  *U \cdot \mbox{\rm{Im}} [L] \cdot U
\end{equation}
where the scalar curvature $R$ and the trace $r$ with the structure
tensor $\Pi$ depend on the scalars $X$ and $Y$ and the $U$
derivatives as:
\begin{equation} \label{R-r-1}
R = 2 (X+Y) + \Phi^2  - \Psi^2 - 2 \delta \Phi \, , \quad \qquad r =
2(X-Y)
\end{equation}

Finally, if we substitute (\ref{re-P}) and (\ref{im-P}) in the
expression (\ref{ricci-2}), we obtain:
\begin{theorem} \label{theo-ricci}
If a space-time admits an umbilical structure $U$, the Ricci tensor
takes the expression:
\begin{equation}
Ric = \frac{1}{4}( \hat{R} \ g + \hat{r} \Pi) + \frac12 ( M - U
\cdot N \cdot *U - *U \cdot N \cdot U)  \label{ricci-3}
\end{equation}
$M$ and $N$ depending on the expansion and rotation vectors $\Phi$
and $\Psi$ as (\ref{M-N}), and
$$
\begin{array}{lll}
2 \hat{R} & \equiv & 4 (X+Y) + 3 \Phi^2 - 3 \Psi^2 - 2 \delta
\Phi\\[1mm]
2 \hat{r} & \equiv & 4 (X-Y) +  \Pi(\Phi, \Phi) - \Pi(\Psi, \Psi) +
2 \delta_{\Pi} \Phi
\end{array}
$$
where $\delta_{\Pi} \Phi \equiv - \Pi^{\alpha \beta} \nabla_{\alpha}
\Phi_{\beta}$.
\end{theorem}
This proposition shows that the Ricci tensor is determined, up to
the scalars $X$ and $Y$ given in (\ref{curvaturas}), by $U$, $\chi$
and ${\cal L}_{\chi} g$.

\subsection{The Weyl tensor}

As a consequence of theorem \ref{theo-integra-con-2}, the self-dual
Weyl tensor ${\cal W}$ is determined by the scalar $\Omega$ and the
self-dual 2--form ${\cal Z} \equiv A_{\perp}$. Indeed, (\ref{i-c-1})
gives the orthogonal components of ${\cal W}$, (\ref{i-c-2}) gives
the mixed components and the ${\cal U} \otimes {\cal U}$ component
is determined by the traceless condition. On the other hand, from
the expression (\ref{T-xi-chi}) of $T$ we can obtain:
\begin{eqnarray}
2 (A,{\cal U}) & = & \delta \chi - \chi^2 \\
{\cal Z}  \equiv  A_{\perp} & = & \frac{1}{2} \dif \xi_{\perp} =
[\nabla \chi , {\cal U} ]_{\perp} = \frac{1}{2} [ \dif \chi , {\cal
U}]_{\perp} = \frac{1}{2} [ \dif \chi , {\cal U}]  \label{zeta}
\end{eqnarray}
and, consequently, we have:
\begin{theorem} \label{theo-weyl}
If a space-time admits an umbilical structure ${\cal U}$, the
self-dual Weyl tensor takes the expression:
\begin{equation} \label{weyl-auto-a}
{\cal W} = 3 \Omega  \, {\cal U} \otimes {\cal U} + \Omega \, {\cal
G} + {\cal U} \stackrel{\sim}{\otimes} {\cal Z}
\end{equation}
where
\begin{equation}
\Omega =  \frac12 (\chi^2 - \delta \chi) - \frac{R}{12} \, , \qquad
{\cal Z} = \frac{1}{2} [ \dif \chi , {\cal U}] \label{weyl-auto-b}
\end{equation}
Moreover, the Weyl tensor invariants are
$$a \equiv  \tr {\cal W}^2 = 2 (3 \Omega^2 - ({\cal Z},{\cal Z})) \, , \qquad b
\equiv \tr {\cal W}^3 = 3 \Omega (({\cal Z},{\cal Z}) - 2 \Omega^2)
$$ and the Weyl tensor eigenvalues
\begin{equation} \label{valpro}
\Omega, \qquad \frac{1}{2} \Big( - \Omega \pm \sqrt{9 \Omega^2 - 4
({\cal Z},{\cal Z})}  \Big)
\end{equation}
\end{theorem}
This theorem shows that the Weyl tensor is determined by the scalar
curvature $R$, and by $U$, $\chi$  and $\dif\chi$. This fact and
theorem \ref{theo-ricci}, which gives the Ricci tensor, show that
the Riemann tensor is determined, up to the scalars $X$ and $Y$, by
the the first and second derivatives of $U$, in accordance with our
comment before expression (\ref{curvaturas}).

From expressions (\ref{weyl-auto-a}) and (\ref{weyl-auto-b}) we can
obtain the real Weyl tensor. Indeed, if we substitute $\chi$ in
terms of the expansion and rotation vectors $\Phi$ and $\Psi$, and
we take the real part of (\ref{weyl-auto-a}), we obtain:
\begin{proposition} \label{pro-weyl-2}
If a space-time admits an umbilical structure $U$ the Weyl tensor
takes the expression:
$$
W= 3 \Omega_1 (U \otimes U - *U \otimes *U) + 3  \Omega_2 \, U
\stackrel{\sim}{\otimes} *U + \Omega_1 \, G + \Omega_2  \, \eta + U
\stackrel{\sim}{\otimes} Z - *U \stackrel{\sim}{\otimes} *Z
$$
where
\begin{eqnarray} \label{Z-real}
Z & = & \frac14 ([\dif \Phi , U] + [\dif \Psi , *U]) ,\\
\Omega_1 & \equiv & \frac{1}{24} ( \Phi^2 - \Psi^2 - 2 \delta \Phi)
- \frac16 (X+Y) \, , \quad \Omega_2 \equiv \frac14 ( (\Phi,\Psi) -
\delta \Psi) \label{v-propi-real-im}
\end{eqnarray}
\end{proposition}

\section{The compatible Petrov-Bel types}
\label{petrov}

The algebraic classification of the Weyl tensor $W$ can be obtained
by studying the traceless linear map defined by the self--dual Weyl
tensor ${\cal W}$ on the SD bivectors space. In terms of the
invariants $a$ and $b$ the characteristic equation reads $\,
x^{3}-\frac{1}{2} ax -\frac{1}{3} b =0  \, $. Then, Petrov-Bel
classification follows taking into account both the eigenvalue
multiplicity and the degree of the minimal polynomial. The
algebraically regular case (type I) occurs when $6b^2 \not= a^3$ and
so the characteristic equation admits three different roots. If
$6b^2 = a^3 \not= 0$, there is a double root and a simple one and
the minimal polynomial distinguishes between types D and II.
Finally, if $a=b=0$ all the roots are equal and so zero, and the
Weyl tensor is of type O, N or III, depending on the degree of the
minimal polynomial. A fully tensorial algorithm has been presented
in \cite{fms} enabling us to determine the Petrov-Bel type.

On the other hand, let us remember that we have defined a Debever SD
bivector as one whose principal directions are Debever principal
null directions of the Weyl tensor \cite{fms}. As the principal
directions of ${\cal U}$ are null shear--free geodesics, they are
Debever principal directions and, consequently, ${\cal U}$ is always
a unitary Debever SD bivector.

In the previous section we have obtained the Weyl tensor of a metric
admitting a 2+2  umbilical  structure as well as the invariants of
the Weyl tensor. These invariants depend on $\Omega$ and $({\cal
Z},{\cal Z})$ as given in theorem \ref{theo-weyl}. Then, a
straightforward calculation leads to:
$$ a^3 - 6 b^2 = - 2 ({\cal Z},{\cal Z})(9 \Omega^2 - 4 ({\cal Z},{\cal Z}))$$
Then, the algorithm given in \cite{fms} works and we can obtain the
algebraic type of the Weyl tensor in terms of ${\cal Z}$ and
$\Omega$:

\begin{theorem} \label{theo-types}
If a space-time admits a 2+2  umbilical structure, then the
canonical SD bivector ${\cal U}$ is a unitary Debever SD bivector.
Thus,
the space-time cannot be of Petrov-Bel type N. Moreover:\\[1mm]
i) It is of type O  if, and only if,  $\ \Omega =0 , \quad {\cal Z}
=0 $.\\[1mm]
ii) It is of type III if, and only if, $\ \Omega =  0 ,
\quad {\cal Z} \neq 0,  \quad ({\cal Z},{\cal Z})  =0$.\\[1mm]
iii) It is of type D  if, and only if, $\ \Omega  \neq 0 , \quad
{\cal Z} =0 $.\\[1mm]
iv) It is of type II  if, and only if, $\ \Omega \neq 0 \neq {\cal
Z}, \quad
({\cal Z},{\cal Z})(9 \Omega^2 - 4 ({\cal Z},{\cal Z})) = 0$.\\[1mm]
v) It is of type I  if, and only if, $\ ({\cal Z},{\cal Z})(9
\Omega^2 - 4 ({\cal Z},{\cal Z})) \neq 0 $.
\end{theorem}

Now we analyze the relationship between the geometry of the Weyl
tensor and the geometrical elements defined by ${\cal U}$ and its
derivatives. We show that both SD bivectors ${\cal U}$ and  ${\cal
Z}$ are closely related to the principal null directions or Debever
directions.

For every Petrov-Bel type, we briefly summarize the geometrical
elements that the Weyl tensor outlines. We use the notation and
results used in the aforementioned paper \cite{fms}, where an
exhaustive approach to this topic can be found. We compare the
canonical form of the Weyl tensor of every type with the expression
of the Weyl tensor obtained in theorem \ref{theo-weyl}.

\subsection{Petrov-Bel type III}

The self-dual Weyl tensor of a type III space-time takes the
canonical expression \cite{fms}:
$${\cal W} = {\cal U} \stackrel{\sim}{\otimes} {\cal H} $$
where ${\cal H}$ is the null eigen-bivector and  ${\cal U}$ the
canonical unitary SD bivector. The two Debever directions are the
principal directions of the canonical SD bivector ${\cal U}$, the
triple one being the common null principal direction to ${\cal H}$
and ${\cal U}$.

From theorem \ref{theo-types}, a type III space-time with an
umbilical structure is characterized by $\Omega =0$  and $({\cal
Z},{\cal Z})= 0$, that is, ${\cal Z}$ is a null SD bivector. Then,
the expression for the Weyl tensor in theorem \ref{theo-weyl}
becomes
$${\cal W} = {\cal U} \stackrel{\sim}{\otimes} {\cal Z} $$
Consequently, we find that ${\cal U}$ coincides with the canonical
unitary SD bivector and ${\cal Z}$ with the null eigen-bivector.
This way, we can conclude
\begin{proposition}
If a type III space-time admits a 2+2 umbilical structure ${\cal
U}$, then the Debever SD bivectors are ${\cal U}$ and ${\cal Z}$,
that is, the Debever directions are the principal directions of
${\cal U}$, the triple one being the fundamental direction of the
null SD bivector ${\cal Z}$.
\end{proposition}

\subsection{Petrov-Bel type D}

Theorem \ref{theo-types} states that a type D space-time admits a
2+2 umbilical structure ${\cal U}$ if, and only if, $\Omega \neq 0$
and ${\cal Z} =0$. Then, expression (\ref{weyl-auto-a}) of the Weyl
tensor becomes:
$$
{\cal W} = 3 \Omega  \, {\cal U} \otimes {\cal U} + \Omega \, {\cal
G}
$$
Thus, ${\cal U}$ is the principal SD bivector of the Weyl tensor
with associated eigenvalue $\Omega$. Then the principal null
directions of ${\cal U}$ are the Debever directions of the Weyl
tensor. Consequently, we conclude:
\begin{proposition}
If a type D space-time admits a 2+2 umbilical structure ${\cal U}$,
then ${\cal U}$ is the principal SD bivector of the Weyl tensor,
that is, the two double Debever directions are the principal
directions of ${\cal U}$.
\end{proposition}

\subsection{Petrov-Bel type II}

A type II Weyl tensor has a unitary eigenbivector  associated with
the simple eigenvalue $- 2 \rho$ and just a null one ${\cal H}$
corresponding to the double eigenvalue $\rho$.  The double Debever
principal null direction $l$ is the fundamental direction of ${\cal
H}$ and, moreover, there exist two simple Debever directions
$l_{\pm}$. Thus, three unitary Debever SD bivectors $\{ {\cal V} ,
{\cal V}_{\pm} \}$ can be considered, ${\cal V}$ having $l_{\pm}$ as
principal directions, and ${\cal V}_{\pm}$ having $l$ and $l_{\pm}$,
respectively, as principal directions. In terms of these Debever SD
bivectors the Weyl tensor takes these two alternative canonical
expressions \cite{fms}:
\begin{equation} \label{tipo2}
{\cal W} = \rho {\cal G} + \frac{3}{2} \rho \ {\cal V}_{+} \stackrel
{\sim}{\otimes} {\cal V}_{-}  = \ - 2 \rho {\cal G} + {\cal H}
\stackrel{\sim}{\otimes} {\cal V}
\end{equation}

As theorem \ref{theo-types} states, type II can occur in two
different ways, when ${\cal Z}$ is a null SD bivector, $({\cal Z},
{\cal Z}) = 0$, or when $9 \Omega^2 - 4 ({\cal Z},{\cal Z}) = 0$,
both $\Omega$ and ${\cal Z}$ being non zero.

In the first case, expressions (\ref{valpro}) give us the double
eigenvalue $\rho = \Omega$. Moreover, from theorem \ref{theo-weyl}
we obtain that ${\cal W} ({\cal Z}) = \Omega {\cal Z}$ and so,
${\cal Z}$ is the canonical null SD bivector. This way ${\cal Z}$
determines the double Debever principal null direction. Moreover, in
this case the expression (\ref{weyl-auto-a}) of the Weyl tensor
becomes:
$$
{\cal W} = \Omega {\cal G} + \frac{3}{2} \Omega \ {\cal U}
\stackrel{\sim}{\otimes} \left[ {\cal U} + \frac{2}{3 \Omega}  {\cal
Z} \right]
$$
By comparing this expression with the first in (\ref{tipo2}) we
conclude that in this case ${\cal U} + \frac {2}{3 \Omega} {\cal Z}$
and ${\cal U}$ are the unitary Debever SD bivectors ${\cal
V}_{\pm}$.

In the second case, ${\cal Z}$ is a non null SD bivector, and
expressions (\ref{valpro}) imply that the double eigenvalue is $\rho
= - \frac{1}{2} \Omega$. Then the SD bivector $\frac{3}{2} \Omega \
{\cal U} + {\cal Z} $ is null, and the expression
(\ref{weyl-auto-a}) of the Weyl tensor can be written as:
$$
{\cal W} - \Omega {\cal G} = {\cal U} \stackrel{\sim}{\otimes} \left
[ \frac{3}{2} \Omega \ {\cal U} + {\cal Z} \right]
$$
Thus, we have $\frac{3}{2} \Omega \ {\cal U} + {\cal Z}  $ coincides
with the ${\cal H}$ of (\ref{tipo2}) and, consequently, it is the
canonical null SD bivector. These results are summarized as follows:
\begin{proposition}
Let ${\cal U}$ be the canonical SD bivector of an umbilical
structure in a type II space-time. It holds:

i) If $({\cal Z}, {\cal Z}) = 0$, then ${\cal Z}$ is the canonical
null SD bivector and thus the fundamental direction of ${\cal Z}$ is
the double Debever direction. This one is also the common principal
directions of the Debever SD bivectors ${\cal U}$ and ${\cal U} +
\frac{2}{3 \Omega} {\cal Z}$. The other two principal directions of
these SD bivectors determine the two simple Debever directions.

ii) If  $9 \Omega^2 - 4 ({\cal Z},{\cal Z}) = 0$, then ${\cal U}$ is
the Debever SD bivector whose principal directions are the two
simple Debever directions, and the double one is the fundamental
direction of the null SD bivector $\frac{3}{2} \Omega \ {\cal U} +
{\cal Z}$
\end{proposition}

\subsection{Petrov-Bel type I}

In an algebraically general space-time the Weyl tensor has three
different eigenvalues $\rho_i$, and an orthonormal frame $\{ {\cal
U}_i \}$ of eigenbivectors of the Weyl tensor can be built. On the
other hand, four simple Debever principal null directions exist that
define six unitary Debever SD bivectors $\{ {\cal V}_{i \pm} \}$. If
we choose a value of the index $i$, say $i=3$, the Weyl tensor can
be written as \cite{fms} \cite{fsI}:
\begin{equation} \label{tipo1}
{\cal W} = \rho_3 \ {\cal G} + \frac{\rho_2 - \rho_1}{2} \ {\cal
V}_{3+} \stackrel{\sim}{\otimes}  {\cal V}_{3-}
\end{equation}

As a consequence of theorem \ref{theo-types}, when the space-time
admits an umbilical structure, the type I case can be characterized
by the scalar condition $({\cal Z},{\cal Z})(9 \Omega^2 - 4 ({\cal
Z},{\cal Z})) \neq 0$ . Then, the expression (\ref{weyl-auto-a}) of
the Weyl tensor can be written as (\ref{tipo1}) by taking:
\begin{eqnarray}
\rho_3 & \equiv & \Omega \, , \qquad \rho_2 -\rho_1  \equiv  2
\sqrt{ 9 \Omega^2 - 4 ({\cal Z},{\cal Z})} \\[2mm]
{\cal V}_{3+} & \equiv & {\cal U} \, ,\qquad  {\cal V}_{3-}  \equiv
 \frac{1}{\sqrt{ 9 \Omega^2 -4 ({\cal Z},{\cal Z})}} ( 3 \Omega
{\cal U} + 2 {\cal Z} )
\end{eqnarray}
Thus, we can state:
\begin{proposition}
Let ${\cal U}$ be an umbilical structure in a type I space-time.
Then the four Debever principal null directions are the principal
directions of the Debever SD bivectors ${\cal U}$ and $3 \Omega
{\cal U} + 2{\cal Z}$.
\end{proposition}

The expression (\ref{tipo1}) is not adapted to the Weyl
eigenbivectors; however, these can be obtained by means of the
projectors given in \cite{fms}. Indeed, using (\ref{weyl-auto-a})
and the expression (\ref{valpro}) of the eigenvalues, we obtain:
\begin{proposition}
In a type I space-time with an umbilical structure ${\cal U}$, the
orthonormal basis of eigenbivectors $\{{\cal U}_i\}$ associated with
the eigenvalues {\rm (\ref{valpro})} are, respectively:
$$
{\cal U}_3 \propto {\cal U} \cdot {\cal Z} \, ,\qquad {\cal U}_1
\propto (3\Omega - \alpha) {\cal U} + 2 {\cal Z} \, , \qquad {\cal
U}_2 \propto (3\Omega + \alpha) {\cal U} + 2 {\cal Z} \, ,
$$
where $\alpha \equiv \sqrt{ 9 \Omega^2 -4 ({\cal Z},{\cal Z})}$.
\end{proposition}

\section{Type D space-times admitting two shear-free geodesic null congruences}
\label{sec-typeD}

In theorem \ref{theo-types} we have shown that the Petrov-Bel type D
space-times admitting two geodesic shear-free null congruences, that
is, an umbilical structure, are those non conformally flat
space-times satisfying ${\cal Z} = 0$. Then, from the expression
$2{\cal Z} = [\dif \chi, {\cal U}]$ (see (\ref{zeta})) and condition
(\ref{maxwellian-chi}) which characterizes a Maxwellian structure,
we obtain the following:
\begin{proposition} \label{pro-max-typeD}
If a non conformally flat space-time admits an umbilical and
Maxwellian structure, then the Weyl tensor is Petrov-Bel type D and
the structure is aligned with the Weyl principal structure.
\end{proposition}
This result has been used elsewhere \cite{fsY} to characterize the
type D space-times admitting a non-null Killing-Yano or conformal
Killing-Yano tensor.

Nevertheless, type D space-times exist with an umbilical and non
Max\-wellian structure. In order to obtain complementary conditions
that guarantee the Maxwellian character of the principal structure
one should consider this lemma:
\begin{lemma}
The complex 2--form $\dif \chi$ admits the following decomposition
in its self-dual and anti-self-dual parts:
\begin{equation} \label{chi-dual-anti}
\dif \chi = ([\dif \chi, {\cal U}] + \{\dif \chi, {\cal U} \})
\cdot{\cal U}
\end{equation}
\end{lemma}
The decomposition (\ref{chi-dual-anti}) is valid for an arbitrary
complex two form, the first term being a self-dual 2--form. The
second term contains, generically, the ${\cal U}$-component but, for
the 2--form $\dif \chi$, this component vanishes as a consequence of
(\ref{max-5}).

From lemma above, theorem \ref{theo-types} and expression
(\ref{zeta}) one finds: {\em a space-time with two shear-free
geodesic null congruences is type D, if and only if, the 2--form
$\dif \chi$ is anti-self dual}, as stated by Dietz and R\"{u}diger
\cite{diru-1}. Then, if this anti-self-dual part vanishes, the
umbilical principal structure of a type D space-time becomes
Maxwellian. But the nullity of a self-dual or anti-self-dual 2--form
is equivalent to the nullity of its real (or imaginary) part. Thus,
we can state:
\begin{proposition}  \label{pro-max-iff}
In a non conformally flat space-time, an umbilical structure ${\cal
U}$ is Maxwellian ($\dif \chi = 0$)
if, and only if, two (and then all) of the following conditions hold:\\[1mm]
(i) The space-time is Petrov-Bel type D (and the structure is the
principal one).\\[1mm]
(ii) The structure is pre-Maxwellian (i.e. the expansion 1--form is
closed, $\ \dif \Phi = 0$).\\[1mm]
(iii) The rotation 1--form is closed, $\ \dif \Psi = 0$.\\[1mm]
(iv) $ \{\dif \chi, {\cal U}\} = 0$ (i.e. $\dif \Phi = *\dif \Psi$).
\end{proposition}

Now we analyze in detail the last condition (iv) in the proposition
above. Under the umbilical condition, the anti-self-dual part of
$\dif \chi$ is $\{S,{\cal U}\}$ as a consequence of the expression
of the tensor $S$ given in theorem \ref{theo-integra-con-2} and the
definition of $\chi$. Then, from the integrability condition
(\ref{i-c-3}), we can state:
\begin{lemma}
Let ${\cal U}$ be the canonical SD bivector of an umbilical
structure and $S$ the tensor defined in theorem
\ref{theo-integra-con-2}. The
following statements are equivalent:\\[1mm]
i) $\dif \chi$ is a self-dual 2-form (i.e. $ \{\dif \chi, {\cal U}\}
= 0$)\\[1mm]
ii) $[ Ric , {\cal U}] = 2S $.
\end{lemma}
As a direct consequence of this lemma and proposition
\ref{pro-max-iff} we obtain:
\begin{corollary} \label{cor-max-ricci}
In a non conformally flat space-time, let ${\cal U}$ be the
canonical SD bivector of an umbilical structure and $S$ the tensor
defined in theorem \ref{theo-integra-con-2}. The
following statements are equivalent:\\[1mm]
i) ${\cal U} $ defines a maxwellian structure.\\[1mm]
ii) The space-time is Petrov-Bel type D and $[ Ric , {\cal U}] =
2S$.
\end{corollary}

Despite this result, the Maxwellian character of an umbilical
structure does not restrict, generically, the algebraic type of the
Ricci tensor. Nevertheless, the Maxwellian condition can restrict
the Ricci tensor if we impose complementary conditions. Thus, we
analyze elsewhere \cite{fsKV} the important role played by tensor
$S$ in generalizing to other energy contents the commutative group
of symmetries that the type D vacuum solutions admit, and we will
show in \cite{fsKV} that in this case the Ricci tensor becomes
algebraically special. In this study corollary \ref{cor-max-ricci}
will play a significative role.

On the other hand, some conditions on the Ricci tensor can impose
strong restrictions on the Weyl principal structure of a type D
space-time. Indeed, we have shown elsewhere \cite{fsD} that, for
vacuum solutions or when the Cotton tensor vanishes, the principal
structure of a type D space-time is umbilical and Maxwellian.
Moreover, in this case, the Weyl eigenvalues are real if, and only
if, the structure is integrable \cite{fsD}. Generalizing this kind
of results to other energy contents is underway and involves
analyzing the Bianchi identities for a type D Weyl tensor and a
Ricci restricted by the umbilical integrability conditions.

Nevertheless, we can obtain some simple results which do not depend
on the Ricci tensor. Thus, from proposition \ref{pro-max-iff} and
the expression (\ref{v-propi-real-im}) of the imaginary part of the
Weyl eigenvalue, we have:
\begin{proposition} \label{pro-in-real}
If a Petrov-Bel type D space-time admits an umbilical and integrable
structure, then the Weyl tensor has real eigenvalues and the
structure is the principal one and Maxwellian.\\[1mm]
Moreover, a function $\phi$ exists such that $\dif \phi = \Phi$ and
the metric tensor is conformal to a product one, the conformal
factor being $e^{-2\phi}$.
\end{proposition}
The second statement of this proposition is a direct consequence of
the changes that a conformal transformation produces on the
differential properties of a structure. A summary of this subject
can be found in \cite{fsD}.

It is worth remarking that all the degenerate static solutions as
well as their charged counterparts satisfy the hypothesis of the
proposition \ref{pro-in-real}. The canonical expression of the
metric as conformal to a product one has allowed us to obtain an
intrinsic and explicit labeling of both families of solutions
\cite{fsD,fsS}.

It has been known for years \cite{hm-1,diru-1} that the two null
eigenvectors of a conformal or a Killing tensor of type 2+2 as well
as the two principal directions of a non-null Killing-Yano or a
conformal Killing-Yano 2--form are shear-free geodesic null
congruences. More recently \cite{fsY,cfsKT} we have recovered this
property and have studied the complementary conditions, as equations
on the structure $U$ defined by these null directions, that
guarantee the existence of these first integrals of the geodesic
equation. For conformal Killing tensors and conformal Killing-Yano
tensors these complementary conditions are related with Maxwellian
properties of the structure. More precisely, we have
\cite{fsY,cfsKT}:
\begin{lemma} \label{lemma-ck-cky}
A 2+2 traceless symmetric tensor is a conformal Killing tensor if,
and only if, it defines an umbilical and pre-Maxwellian
structure.\\[1mm]
A non-null 2--form is a conformal Killing-Yano tensor if, and only
if, it defines an umbilical and Maxwellian structure.
\end{lemma}

It is known that the type D vacuum solutions which admit a Killing
tensor also admit a Killing-Yano tensor \cite{collinson,ste}, and
all the type D vacuum solutions admit a conformal Killing tensor
\cite{walker-p}. The last statement can be generalized to type D
solutions with a vanishing Cotton tensor and extended to conformal
Killing-Yano tensors as a consequence of the results in \cite{fsD}
and lemma \ref{lemma-ck-cky}. Moreover, from this lemma and
proposition \ref{pro-max-iff} we obtain an interesting result which
does not depend on the energy content:
\begin{theorem}
Every Petrov-Bel type D space-time that admits a 2+2 conformal
Killing tensor also admits a conformal Killing-Yano tensor.
\end{theorem}

{\bf Acknowledgements} This work has been partially supported by the
Spanish Ministerio de Educaci\'on y Ciencia, MEC-FEDER project
FIS2006-06062.

\end{document}